\documentclass[journal]{IEEEtran}

\usepackage{cite}

\usepackage{graphicx}
                        
\usepackage{hyperref}
\usepackage{amsmath}

\begin{document}

\title{Interrogation of a fiber Fabry-Perot sensor by current modulation of a diode laser}

\author{Jong H. Chow,
        Jeff S. Cumpston,
        Ian C.M. Littler\dag,
        David E. McClelland,
        and~Malcolm~B.~Gray% <-this % stops a space
\thanks{This research was supported by the Australian Research Council (ARC) under the auspices of the Australian Consortium for Interferometric Gravitational Astronomy, with partial assistance from the Centre for Ultrahigh Bandwidth Devices for Optical Systems (CUDOS).}%
\thanks{The authors are with the Centre for Gravitational Physics, Faculty of Science, The Australian National University, Canberra, ACT 0200, Australia; \dag Ian C.M. Littler is with CUDOS, School of Physics, A28, University of Sydney, Camperdown, NSW 2006, Australia.}}

\markboth{Journal of IEEE Photonics Technology Letters,~Vol.~, No.~,~}{Shell \MakeLowercase{\textit{Jong H. Chow et al.}}: Signal extraction by current modulation of a diode laser in a fiber resonator based sensing system}

\maketitle

\hfill{\today}

\begin{abstract}
We present a method for remote interrogation of passive fiber Bragg grating Fabry-Perot resonators, employing current modulation of the diode laser source.  With the presence of both RF amplitude and frequency modulation, it is a variant of the ultra-sensitive Pound-Drever-Hall laser frequency locking technique.  We demonstrate that current modulation and interferometric demodulation removes the need for a phase modulator in the sensing architecture.
\end{abstract}

\begin{keywords}
fiber sensor, modulation, interferometry, strain sensing, fiber resonator.
\end{keywords}

\IEEEpeerreviewmaketitle

%\section{Introduction}

\PARstart{T}{he} Pound-Drever-Hall frequency (PDH) locking technique \cite{Drever, Black} is widely used in the gravitational wave detection community for a range of applications, including laser frequency stabilization \cite{Day}, interferometer longitudinal and alignment control, as well as gravitational wave signal extraction \cite{Strain, Shaddock}.  PDH locking is the method of choice for ultra-sensitive interferometry.  While it is well-established with free-space bulk-optical resonators and solid-state lasers, it can be readily extended to diode laser stabilization and guided-wave optics.  In fiber optics, Erbium doped fiber laser frequency stabilization using PDH locking has been demonstrated \cite{Park}.  

We have recently shown that the PDH technique can also be used for signal extraction in a fiber Bragg grating Fabry-Perot (FFP) sensor, achieving frequency noise performance compatible with pico-strain resolution \cite{Chow}.  This is an extremely powerful technique for both static and dynamic strain sensing because the PDH system is immune to laser intensity noise to the first order when the laser is locked to a resonator.

In this Letter we present results for a modified PDH locking technique, where we generate RF sidebands by directly modulating the drive current of the interrogating laser.  We compare its fiber sensing performance with that obtained using pure phase modulation (PM).  Direct current modulation has the advantage of removing the need for a phase modulator in the sensing architecture.  This offers a significant cost saving and simplifies the packaging of the sensing system.  

\begin{figure}[htb]
\centering
\includegraphics[width=3.0in]{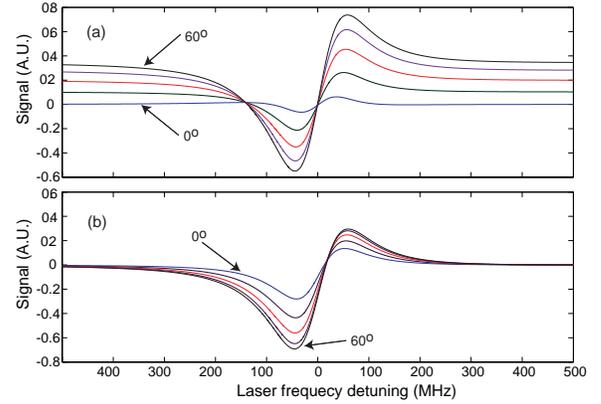}
\caption{Theoretical plots for (a) reflected and (b) transmitted error signals for various demodulation phase shift $\Delta\psi$ in 15$^{o}$ steps, when an FFP of 150MHz linewidth is interrogated with current modulation sidebands of 15MHz.  Zero detuning in the figures corresponds to FFP resonance center.}
\label{theo_errsig}
\end{figure}

Current modulation of a diode laser introduces both amplitude modulation (AM) and frequency modulation (FM).  This can change the symmetry and zero-crossing of the demodulated error signal.  For small modulation depths, the electric field of the incident laser light can be written as:
\begin{eqnarray}
  \tilde{E}_{inc} & = & E_{0}[1+\alpha \cos(2 \pi \nu_{m})]
  e^{i{[2 \pi \nu -\delta \cos(2 \pi \nu_{m} t)]t}}                    \nonumber        \\ 
  & \approx & E_{0}[1 + \frac{1}{2} (\alpha + \beta) e^{i 2 \pi \nu_{m}t}    \nonumber  \\
  & & + \frac{1}{2} (\alpha - \beta) e^{-i 2 \pi \nu_{m}t}],
  \label{inc_field}
\end{eqnarray}
where ${E}_{0}$ is the DC field amplitude; $\nu$ is the optical carrier frequency; $\nu_{m}$ is the modulation frequency; $\alpha$ is the effective AM modulation depth; $\delta$ is the effective FM modulation depth while $\beta$ is its PM equivalent; and $\alpha$, $\delta$ and $\beta$ are all $\ll$ 1.  We can see from equation \ref{inc_field} that the incident field in the presence of both AM and FM can be represented by a carrier with two sidebands of unequal amplitudes.  When $\alpha = \beta$, the input field becomes the special case for single sideband modulation \cite{Cusack}.

The current modulated laser is used to interrogate an FFP, while measuring either the reflected or transmitted power with a photodetector.  We derive the demodulated and low-pass filtered signal as
\begin{eqnarray}
  V(\nu) &\propto& \Re[\tilde{F}(\nu)\tilde{F}^{*}(\nu_{+})(\beta+\alpha)  \nonumber \\
               & & -\tilde{F}^{*}(\nu)\tilde{F}(\nu_{-})(\beta-\alpha)]\cos(\Delta\psi) \nonumber \\
               & & +\Im[\tilde{F}(\nu)\tilde{F}^{*}(\nu_{+})(\beta+\alpha)  \nonumber \\
               & & -\tilde{F}^{*}(\nu)\tilde{F}(\nu_{-})(\beta-\alpha)]\sin(\Delta\psi),
  \label{err_v}
\end{eqnarray}
where $\tilde{F}(\nu)$ and $\tilde{F}^{*}(\nu)$ are the complex response of the FFP and its conjugate, either in transmission or reflection, at frequency $\nu$; $\nu_{+} = \nu + \nu_{m}$ and $\nu_{-} = \nu - \nu_{m}$; while $\Delta\psi$ is the adjustable RF phase shift to choose the demodulation quadrature.

Figs. \ref{theo_errsig}(a) and \ref{theo_errsig}(b) illustrate the demodulated error signals for reflection and transmission respectively, using equation \ref{err_v}, for various demodulation phase $\Delta\psi$ in 15$^{o}$ steps.  We have assumed that the grating pair which make up the FFP are of equal reflectivity, and the single pass loss in the resonator is 1$\%$ of the circulating power.  The FFP resonance linewidth used for the calculations is 150MHz, while the current modulation frequency is 15MHz.  The $\beta$:$\alpha$ ratio assumed in the plots is 50:1.

% to approximate the current modulation of the laser we used in the experiment, which was effectively dominated by PM.

It can be seen from Fig. \ref{theo_errsig}(a) that the error signals differ from classic PDH error signals in two ways.  They have a DC offset when the laser carrier is off-resonance because of the unequal amplitudes of the sidebands.  Also, there is a slight deviation from the anti-symmetric shape of a classic PDH error signal.  The zero-crossing in the error signals, however, are very close to resonance, which is the same as PDH error signals with pure phase modulation.  The zero-crossing can be expected to intersect exactly on resonance when the FFP is impedance matched, ie, the two reflectors are matched and the FFP is lossless.  In an impedance matched resonator, no light is reflected on resonance, implying that $\tilde{F}(\nu)=0$, and hence equation \ref{err_v} becomes zero.

Fig. \ref{theo_errsig}(b), on the other hand, shows that transmitted error signals tend to zero off-resonance, because there is very little carrier or sideband transmitted outside resonance.  When the laser is near resonance, the transmitted error signals exhibit even more pronounced asymmetry compared to the reflection case.  For both cases, this asymmetry is enhanced as the AM component of the modulation increases, ie. the $\beta$:$\alpha$ ratio decreases.  More importantly, in transmission the zero-crossing in the error signal is detuned slightly from the FFP resonance, where the amount of detuning depends on the demodulation phase $\Delta\psi$.  For a given resonator linewidth and modulation frequency, this detuning is greater as the $\beta$:$\alpha$ ratio is reduced.

\begin{figure}[htb]
\centering
\includegraphics[width=3.0in]{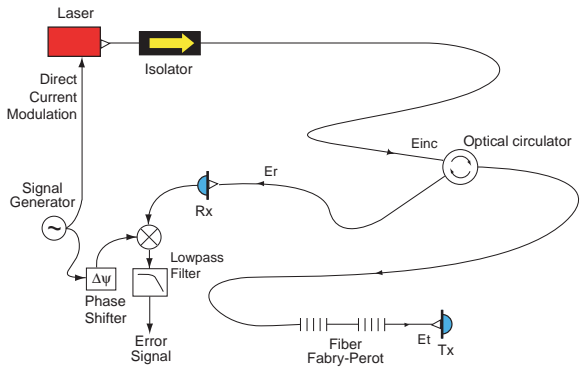}
\caption{The experimental schematic of the current modulated PDH interrogation sensing system.  $E_{inc}$, $E_{r}$ and $E_{t}$ are the input, reflected and transmitted electric fields of the laser, respectively.  Rx is the reflection photodetector, while Tx is for the transmission.}
\label{schematic}
\end{figure}

This implies that in frequency locking applications that require the laser to be locked to line-center, for maximum circulating and transmitted power, reflection locking is the preferred mode of operation.  As for sensing purposes, the exact lock point relative to resonance center is less crucial, provided the slope of the error signal around the lock point is sufficiently linear.  Thus either reflection or transmission modes can be utilized, depending on the preferred sensing architecture.

The experimental setup for the current RF modulated interrogation scheme is illustrated in Fig. \ref{schematic}.  The laser was a New Focus Vortex 6029, which is a tunable extra-cavity diode laser centered at 1550.15nm, with about 0.40nm, or $\sim$50GHz tuning range.  The frequency tuning of the laser was actuated by applying a voltage to the piezo-electric transducer (PZT) of its cavity.  Our FFP consisted of a pair of nominally matched 13.5dB Bragg gratings (R $\approx$ 95.5$\%$) each 15mm long, spaced 10mm apart.  The selected resonance for this study had a linewidth of approximately 143MHz.  The current modulation was driven by an RF signal generator at 15MHz, which also provided the local oscillator for the demodulation electronics.  The local oscillator was phase shifted by $\Delta\psi$ before being used to mix down the electronic signal from either the reflected or transmitted port.  This mixed output was then low-pass filtered to provide the demodulated error signal.  

\begin{figure}[htb]
\centering
\includegraphics[width=3.0in]{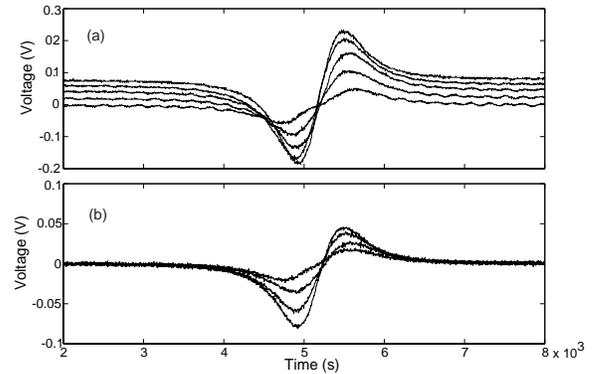}
\caption{The experimental current modulation error signal traces for (a) reflection, and (b) transmission as the laser frequency was swept through the resonance of a fiber Fabry-Perot.  Each shows overlaid error signals for several RF phase shifts $\Delta\psi$.}
\label{exp_errsig}
\end{figure}

To observe the error signals in the presence of both AM and FM, the frequency of the laser was scanned by applying a 50:50 sawtooth from a function generator.  The error signals were recorded with a digital oscilloscope and displayed in Fig. \ref{exp_errsig}.  Fig. \ref{exp_errsig}a plots the family of reflected error signals as the RF phase shift $\Delta\psi$ was varied.  They show only slight asymmetry with differing off-resonance DC offset, as predicted by Fig. \ref{theo_errsig}a.  It can be seen that they intersect very near zero-crossing, as our FFP was nearly impedance matched.

Similarly, Fig. \ref{exp_errsig}b presents the error signals in transmission.  As predicted in Fig. \ref{theo_errsig}b, the off-resonance DC offset seen in reflection was absent in the transmission scans, and their asymmetry was stronger compared with the former.  The agreement in the amount of asymmetry between the experimental and theoretical error signals, as well as the small amount of zero-crossing detuning in transmission, would indicate that current modulation in our laser is dominated by PM, and the $\beta$:$\alpha$ ratio is $\simeq$ 50.

\begin{figure}[htb]
\centering
\includegraphics[width=3.0in]{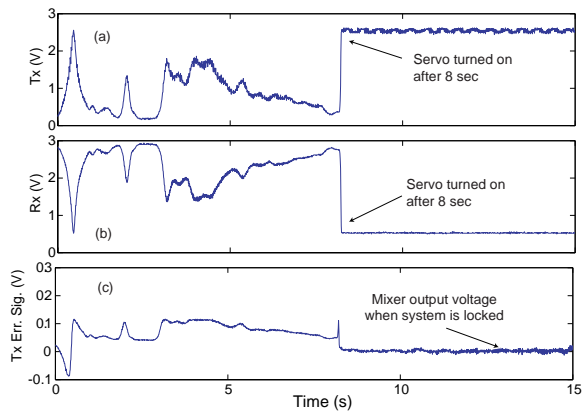}
\caption{The experimental traces for (a) transmitted intensity; (b) reflected intensity; and (c) transmitted error signal from the mixer output during lock acquisition.}
\label{exp_lock_acq}
\end{figure}

To lock the laser to the FFP resonance, we first selected the largest error signal available in the transmission (see Fig. \ref{exp_errsig}b) by adjusting $\Delta\psi$, after which the sawtooth to the PZT was replaced with the amplifed error signal.  Before the feedback loop was closed, the PZT DC offset voltage was tuned slowly while the transmitted and reflected laser intensities were monitored.  When the laser nears resonance, the transmitted intensity approaches its maximum, and the feedback loop was then engaged to acquire lock.  This process was recorded with a digital oscilloscope, and the traces are illustrated in Fig. \ref{exp_lock_acq}.  The servo amplifier used in this experiment had a single real pole response with a corner frequency of 0.03Hz.  The total feedback loop had a DC gain of $\approx$1000 and a unity gain bandwidth of around 40Hz.  Lock aquisition was both straight forward and robust.  Like the traditional PDH interrogation method, once locked, the laser typically stayed locked for over 3 hours \cite{Chow}.  This duration is limited by the long term FFP thermal drift.

The signal extraction from an FFP sensor, when interrogated with a current modulated laser, is similar to the PDH locking scheme.  This is done by either monitoring the PZT feedback voltage required to maintain lock within the servo unity gain bandwidth, or by monitoring the mixer output at frequencies above the unity gain bandwidth.  Environmental stimulations, such as temperature drift as well as stress and strain due to mechanical or acoustic perturbation, change the resonance condition of the FFP.  These result in both DC and AC voltage change in the mixer output and the PZT feedback voltage.  When the mixer output is read with a dynamic signal analyzer, frequency domain information about these perturbations can be extracted.  The signal analyzer performs a Fast Fourier Transform of the time domain mixer output voltage (shown in Fig. \ref{exp_lock_acq}c), and when the slope of the error signal is used for calibration, it provides a direct measurement of both the frequency noise of the laser source and the dynamic strain in the FFP.

\begin{figure}[htb]
\centering
\includegraphics[width=3.0in]{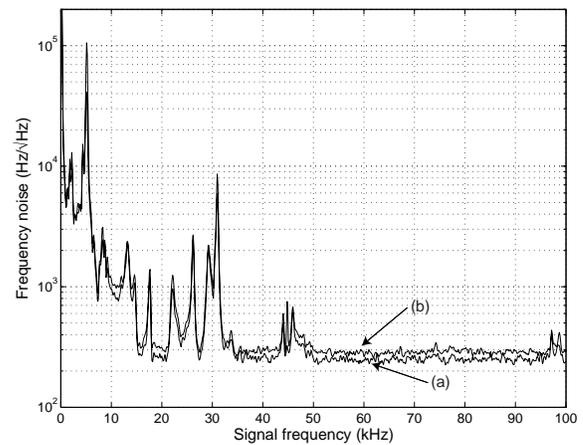}
\caption{The calibrated frequency noise as measured by a dynamic signal analyzer, when a fiber Fabry-Perot was interrogated with a diode laser that was (a) purely phase modulated; and (b) current modulated.}
\label{exp_PSD}
\end{figure}

We compared the performance of the FFP sensor when the laser was purely phase modulated, with the case of current modulation.  The result is illustrated in Fig. \ref{exp_PSD}, where we overlaid the FFT measurements for both cases, and the chosen interrogation architecture was in transmission.  It can be seen that both methods showed the same components of audio frequency ambient noise, including the PZT resonance due to closed-loop excitation, and broadband acoustic noise at low frequencies.  The two modulation schemes yielded comparable broadband sensitivities, with only a slight penalty in adopting current modulation.

In summary, direct modulation of the drive current in our diode laser introduces both AM and FM in its electric field.  We derived the theoretical expression for the expected error signals when an FFP is interrogated with a current modulated diode laser, and they agreed well with experimental data.  We compared the current modulated sensing architecture with that utilizing pure phase modulation, and found that they had comparable sensitivities.  This is an important result, as it eliminates the need for a phase modulator in the sensing topology, with little penalty in observed performance.

%\subsection{Subsection Heading Here}
%Subsection text here.

% needed in second column of first page if using \pubid
%\pubidadjcol

%\subsubsection{Subsubsection Heading Here}
%Subsubsection text here.

% Reminder: the "draftcls" or "draftclsnofoot", not "draft", class option
% should be used if it is desired that the figures are to be displayed while
% in draft mode.

% An example of a floating figure using the graphicx package.
% Note that \label must occur AFTER (or within) \caption.
% For figures, \caption should occur after the \includegraphics.
%
%\begin{figure}
%\centering
%\includegraphics[width=2.5in]{myfigure}
% where an .eps filename suffix will be assumed under latex, 
% and a .pdf suffix will be assumed for pdflatex
%\caption{Simulation Results}
%\label{fig_sim}
%\end{figure}

% An example of a double column floating figure using two subfigures.
% (The subfigure.sty package must be loaded for this to work.)
% The subfigure \label commands are set within each subfigure command, the
% \label for the overall fgure must come after \caption.
% \hfil must be used as a separator to get equal spacing
%
%\begin{figure*}
%\centerline{\subfigure[Case I]{\includegraphics[width=2.5in]{subfigcase1}
% where an .eps filename suffix will be assumed under latex, 
% and a .pdf suffix will be assumed for pdflatex
%\label{fig_first_case}}
%\hfil
%\subfigure[Case II]{\includegraphics[width=2.5in]{subfigcase2}
% where an .eps filename suffix will be assumed under latex, 
% and a .pdf suffix will be assumed for pdflatex
%\label{fig_second_case}}}
%\caption{Simulation results}
%\label{fig_sim}
%\end{figure*}


\begin{thebibliography}{1}

\bibitem{Drever} R. W. P. Drever, J. L. Hall, F. V. Kowalski, J. Hough, G. M. Ford, A. J. Munley, and H. Ward, "Laser phase and frequency stabilization using an optical resonator," \textit{Appl. Phys. B}, vol. 31, pp. 97-105, 1983.

\bibitem{Black} Eric D. Black, "An introduction to Pound-Drver-Hall laser frequency stabilization," \textit{Am. J. Phys.}, vol 69, pp. 79-87, 2001.

\bibitem{Day} Timothy Day, Eric K. Gustafson, and Robert L. Byer, "Sub-Hertz relative frequency stabilization of two-diode laser-pumped Nd:YAG lasers locked to a Fabry-Perot interferometer," \textit{IEEE J. Quantum Electron.}, vol. 28, pp. 1106-1117, 1992.

\bibitem{Strain} Kenneth A. Strain, Guido M\"{u}ller, Tom Delker, David H. Reitze, David B. Tanner, James E. Mason, Phil A. Willems, Daniel A. Shaddock, Malcolm B. Gray, Conor Mow-Lowry, and David E. McClelland, "Sensing and Control in dual-recycling laser interferometer gravitational-wave detectors," \textit{Appl. Opt.}, vol. 42, pp. 1244-1256, 2003.

\bibitem{Shaddock} Daniel A. Shaddock, Malcolm B. Gray, Conor Mow-Lowry, and David E. McClelland, "Power-recycled Michelson interferometer with resonant sideband extraction," \textit{Appl. Opt.}, vol. 42, pp. 1283-1295, 2003.

\bibitem{Park} Namkyoo Park, Jay W. Dawson, and Kerry J. Vahala, "Frequency locking of an erbium-doped fiber ring laser to an external fiber Fabry-Perot resonator," \textit{Opt. Lett.}, vol. 18, pp. 879-881, 1993.

\bibitem{Chow} Jong H. Chow, Ian C. M. Littler, Glenn de Vine, David E. McClelland, and Malcolm B. Gray, "Phase-sensitive interrogation of fiber Bragg grating resonators for sensing applications," Submitted to J. Lightwave Technol.

\bibitem{Cusack} Benedict J. Cusack, Benjamin S. Sheard, Daniel A. Shaddock, Malcolm B. Gray, Ping Koy Lam, and Stan E. Whitcomb, "Electro-optic modulator capable of generating simultaneous amplitude and phase modulation," \textit{Appl. Opt.}, vol. 43, pp. 5079-5091, 2004.

\end{thebibliography}
\end{document}